%% file: main.tex
\documentclass[10pt,twocolumn,letterpaper]{article}

\usepackage[pagenumbers]{cvpr} % To force page numbers, e.g. for an arXiv version

\input{preamble}

\definecolor{cvprblue}{rgb}{0.21,0.49,0.74}
\usepackage[pagebackref,breaklinks,colorlinks,allcolors=cvprblue]{hyperref}

\def\paperID{3712} % *** Enter the Paper ID here
\def\confName{CVPR}
\def\confYear{2025}

\title{M3DA: Benchmark for Unsupervised Domain Adaptation in 3D Medical Image Segmentation}
\author{Boris Shirokikh\\
AIRI, Skoltech\\
{\tt\small shirokikh@airi.net}
\and
Anvar Kurmukov\\
AUMI.AI\\
\and
Mariia Donskova\\
IITP RAS\\
\and
Valentin Samokhin\\
IITP RAS, AIRI\\
\and
Mikhail Belyaev\\
AUMI.AI\\
\and
Ivan Oseledets\\
AIRI, Skoltech\\
{\tt\small oseledets@airi.net}
}

\begin{document}
\maketitle

\input{content/0_abstract}

\input{content/1_intro}

\input{content/2_benchmark}

\input{content/3_methods}

\input{content/4_experiments}
\input{content/5_discussion}

{
    \small
    \bibliographystyle{ieeenat_fullname}
    \bibliography{main}
}

\input{content/X_suppl}

\end{document}

%% file: preamble.tex
\usepackage[utf8]{inputenc} % allow utf-8 input
\usepackage[T1]{fontenc}    % use 8-bit T1 fonts
\usepackage{url}            % simple URL typesetting
\usepackage{booktabs}       % professional-quality tables
\usepackage{amsfonts}       % blackboard math symbols
\usepackage{nicefrac}       % compact symbols for 1/2, etc.
\usepackage{microtype}      % microtypography
\usepackage{xcolor}         % colors
\usepackage{array}
\usepackage{arydshln}
\usepackage{color}
\usepackage{makecell}
\usepackage{multirow}
\usepackage{tabularx}
\usepackage{tcolorbox}
\usepackage{transparent}
\usepackage{xcolor}
\usepackage{amsmath}
\usepackage{wrapfig,lipsum}
\usepackage{tipa}
\usepackage{pifont}         % adds nice "v" and "x" symbols
\usepackage[normalem]{ulem}
\usepackage{tikz}
\usetikzlibrary{arrows,positioning,fit,backgrounds,shapes.geometric,calc,bending}
\usepackage{amsmath}
\usetikzlibrary{arrows,positioning,fit,backgrounds,shapes.geometric,calc}

\newcommand{\ra}[0]{\rightarrow}
\newcommand{\R}[0]{\mathbb{R}}

\newcommand{\cmark}{\ding{51}}%
\newcommand{\xmark}{\ding{55}}%

%% file: content/0_abstract.tex
% Проблема: есть доменный сдвиг
% Актуальность: нет бенчмарка
% Решение: предлагаем системный бенчмарк
% Новизна: МРТ, КТ, первыми системно затрагиваем все ключевые сдвиги
% Ценность: бенчмарк не решается с нахрапа - нужно долбить методы

\begin{abstract}

Domain shift presents a significant challenge in applying Deep Learning to the segmentation of 3D medical images from sources like Magnetic Resonance Imaging (MRI) and Computed Tomography (CT). Although numerous Domain Adaptation methods have been developed to address this issue, they are often evaluated under impractical data shift scenarios. Specifically, the medical imaging datasets used are often either private, too small for robust training and evaluation, or limited to single or synthetic tasks.

To overcome these limitations, we introduce a M3DA \textipa{/"mEd@/} benchmark comprising four publicly available, multiclass segmentation datasets. We have designed eight domain pairs featuring diverse and practically relevant distribution shifts. These include inter-modality shifts between MRI and CT and intra-modality shifts among various MRI acquisition parameters, different CT radiation doses, and presence/absence of contrast enhancement in images.

Within the proposed benchmark, we evaluate more than ten existing domain adaptation methods. Our results show that none of them can consistently close the performance gap between the domains. For instance, the most effective method reduces the performance gap by about 62\% across the tasks. This highlights the need for developing novel domain adaptation algorithms to enhance the robustness and scalability of deep learning models in medical imaging.

We made our M3DA benchmark publicly available: \href{https://github.com/BorisShirokikh/M3DA}{https://github.com/BorisShirokikh/M3DA}.

\end{abstract}

%% file: content/1_intro.tex
\section{Introduction}
\label{sec:intro}

\begin{figure}
    \includegraphics[width=\linewidth]{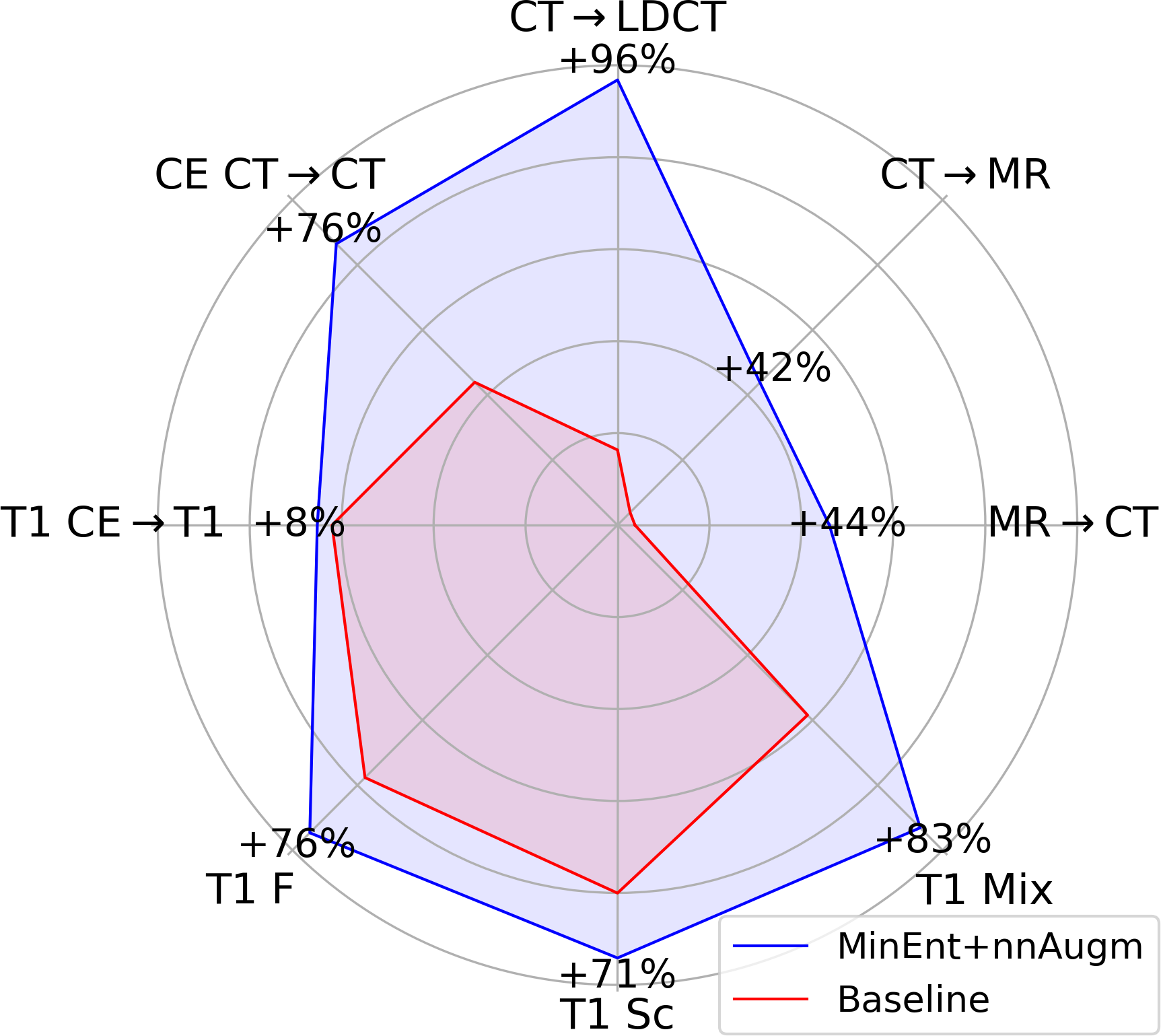}
    \caption{Using the best DA method in the M3DA benchmark closes only 62\% of the performance gap between domains on average. Here, \% indicates the gap closed between the baseline level and oracle (outer) circle.}
    \label{fig:teaser1}
\end{figure}

\begin{figure*}[ht]
    \centering
    \includegraphics[width=1\linewidth]{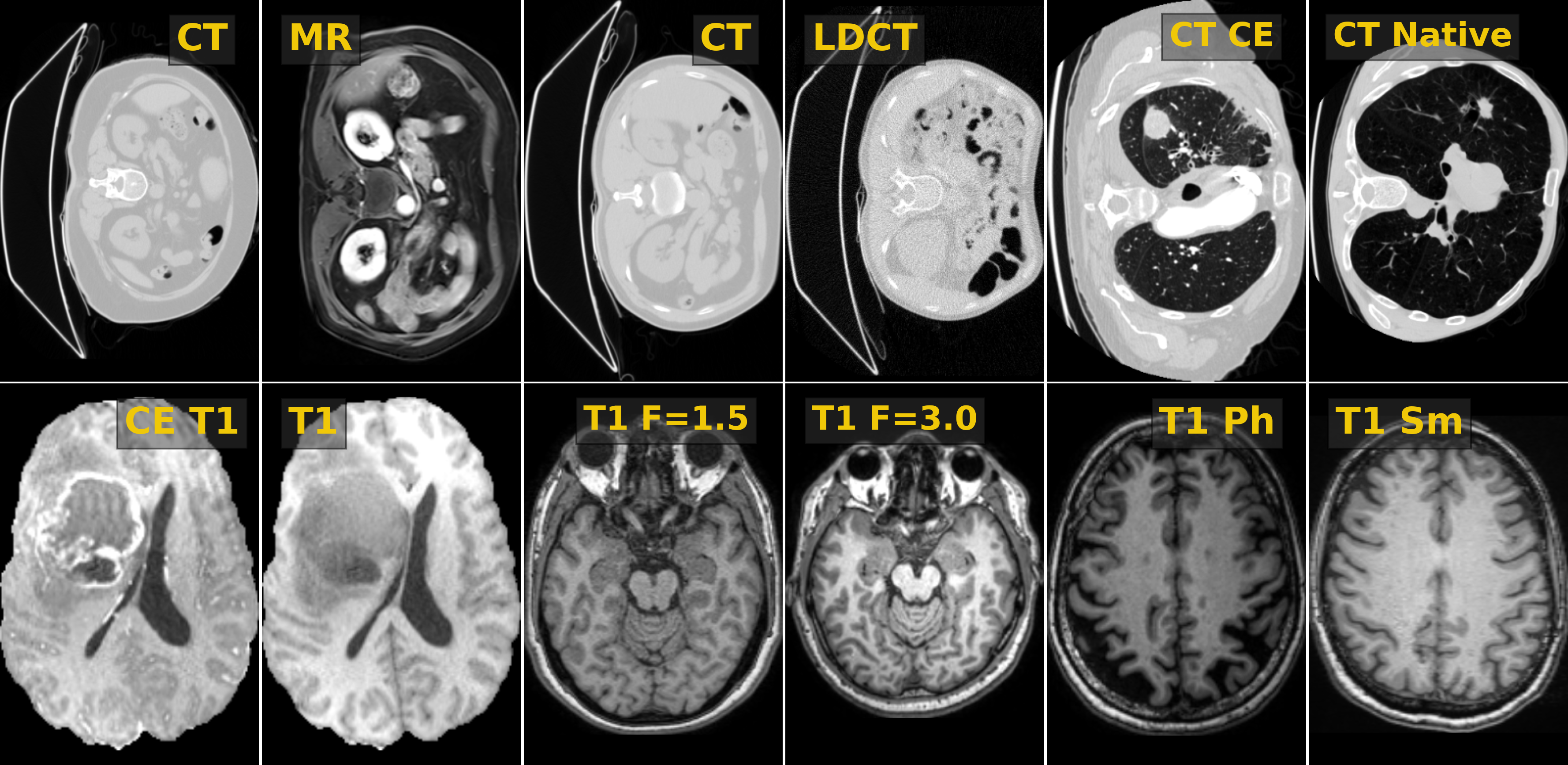}
    \caption{Examples from individual domains in M3DA without segmentation masks for visual comparison between domains. Left to right, top to bottom: CT to MR, CT to LDCT, CT CE to CT native, CE T1 to T1, T1 Field (1.5T to 3T), T1 Scanner (Philips to Siemens). We provide segmentation masks visualization for the same examples in Supplementary materials.}
    \label{fig:teaser2}
\end{figure*}

Deep Learning (DL) methods have significantly advanced medical image analysis, achieving near-human-level performance in tasks like image classification, segmentation, pathology detection, and automated diagnosis \cite{medim_dl_survey_2017}. However, the widespread adoption of DL in medical imaging is hindered by the poor performance of neural networks on data from distributions different from their training set. This challenge, known as \textbf{domain shift}, is particularly prevalent in medical imaging due to changes in scanner acquisition parameters, the introduction of new imaging modalities, and population differences. Several studies have concluded that medical imaging is a crucial domain for the adoption of Domain Adaptation (DA) methods \cite{gulrajani2020search,uda_survey_2020,zhuang2020comprehensive,peng2018visda,zhang2021empirical}. Despite this, there is a lack of a common benchmark for testing DL methods in the field of 3D medical imaging.

% TODO: move fig 1 here, after finishing todo-list

Recent works have focused on developing DA methods, revealing that 3D medical image segmentation algorithms are particularly susceptible to domain shift. Researchers have tested their methods against various sources of domain shift, including shifts between imaging modalities, the most common being MRI and CT \cite{jiang2020unified,yu2023source,zheng2021hierarchical}. Other sources of domain shift include scanner manufacturers or settings, such as the strength of MR field or CT dosage \cite{zheng2021hierarchical,liu2020shape,chen2022maxstyle,gu2021domain,lennartz2023segmentation,se_medim}, and intra-modality shifts, such as T2 to T1 MRI \cite{han2021deep,crossmoda,dann_medim}.

A systematic comparison of these methods, along with the question of the necessity to develop new ones, is complicated by lack of consistency in the usage of datasets across studies, even when addressing similar domain shifts, as shown in Table \ref{tab:benchmarks}. In addition, many studies test their proposals on a single domain shift problem, limiting the generalizability of their analysis.

Recently, the Cross-Modality Domain Adaptation (CrossMoDA) challenge \cite{crossmoda}, conducted at MICCAI in 2021 and repeated in 2022 and 2023, attempted to unify various authors under the same framework. The challenge's setup involved training models on annotated MR T1c with access to unannotated T2 studies, followed by testing on T2 studies. The challenge attracted numerous participants, with the top teams achieving near-supervised performance levels using image-to-image translation techniques paired with multi-stage pseudo-labeling.

Despite its success, the CrossMoDA challenge has certain limitations. Firstly, it only considers a single source of domain shift, the MR sequence. Secondly, the segmentation task is confined to a two-class challenge: segmenting the vestibular schwannoma tumor and the cochlea, both located in very specific anatomical regions. Consequently, top-performing solutions managed to close up to a 97\% score gap between domains. Participants employed various task-specific techniques, such as using only the largest connected component of the segmentation mask to enhance segmentation quality. While this approach may be effective for solving a concrete segmentation task, it is counterproductive for assessing the capabilities of DA methods. It obscures the true impact of adaptation to the unseen domain, hindering a clear understanding of the effectiveness of DA techniques.

\textit{Therefore, we conclude that there is a need for a large, diverse, and publicly available benchmark for DA in 3D medical image segmentation that includes a variety of downstream tasks.} We introduce such a benchmark to encourage further progress in developing scalable DA methods.

We include four publicly available datasets, encompassing 22 segmentation tasks. Based on these datasets, we construct eight domain adaptation problems; see visual examples of individual domains in Figure~\ref{fig:teaser2}. Table~\ref{tab:setup} summarizes all proposed problems with their domain shifts and dataset splits.

To establish a baseline and determine which problems remain unsolved with current methods, we implemented core unsupervised DA (UDA) methods for 3D medical image segmentation. Excluding any task-specific assumptions (e.g., filtering the largest connected component) and human-in-the-loop approaches, the best method within our benchmark closes only 62\% of the performance drop between domains on average. Thus, we highlight the need for further development of robust DA methods in 3D medical image segmentation, with our proposed benchmark serving as a strong foundational point for systematically comparing novel methods. We summarize our contributions as follows:

\begin{itemize}

\item We propose a benchmark for DA in 3D medical image segmentation that includes eight carefully selected domain shifts based on their practical relevance. These shifts cover variations in imaging modalities, scanner settings, and the presence of contrast agents, ensuring that our benchmark reflects real-world challenges in medical image analysis. % (MRI to CT)

\item We provide a comprehensive evaluation of more than ten core domain adaptation methods on our benchmark, covering key categories of UDA approaches. 

\item Our benchmark is designed to be economical, utilizing only four publicly available datasets. This allows for testing new methods against a wide variety of problems with minimal resources, making our benchmark accessible to researchers and encouraging wider adoption.
    
\end{itemize}

%% file: content/2_benchmark.tex
\section{M3DA Benchmark}
\label{sec:bench}

We consider a semantic segmentation problem of 3D medical images, which we call a downstream task. Any downstream model works with input samples $x \in X$ and the corresponding segmentation masks $y \in Y$, where $X$ and $Y$ are some input image and label spaces. If $x \in \R^{H \times W \times D}$, segmentation mask is of the same spatial size $y \in \R^{H \times W \times D}$, where every element belongs to a predefined set of labels $y^{(h,w,d)} \in \{ 0, 1, \dots, C \}$, $0$ is background and $C$ is the number of foreground classes.

\input{tables/1_benchmarks}

We follow the standard unsupervised domain adaptation (UDA) problem setting, as in \cite{dann}. We assume that two distributions $\mathcal{S}(x, y)$ and $\mathcal{T}(x, y)$ exist on $X \otimes Y$, called \textit{source} and \textit{target} distributions. At the training time, we have a set of source training samples $X^s = \{ x_i^s \}_{i=1}^n$ with the corresponding masks $Y^s = \{ y_i^s \}_{i=1}^n$ and a set of target training samples $X^t_{tr}$ without annotations; source images and masks are considered to be sampled from $\mathcal{S}$, $(x_i^s, y_i^s) \sim \mathcal{S}(x, y)$. Our goal is to predict segmentations $y$ given the input from the marginal distribution of target images, $x \sim \mathcal{T}(x)$. To evaluate algorithms, we have target testing samples $X^t_{ts}$ with masks $Y^t_{ts}$ \textit{available only for evaluation purposes}.

Thus, given domains A and B, one trains a supervised model on domain A while having access to unannotated samples from domain B for adaptation. The goal of UDA is to develop a model that makes accurate predictions on domain B. Importantly, this setup prohibits incorporating annotations from the target domain into the training routine.

\subsection{UDA problems motivation}
\label{ssec:constructing}

\input{tables/2_datasets}

\paragraph{CT \textbf{$\leftrightarrow$} MRI}
First, we include domain shift from MRI to CT and vice versa. Although the use of CT scans is often clinically justified, it is associated with additional risks, such as potentially increasing the risk of cancer \cite{cao2022ct,brenner2007computed}. In contrast, MRI is a safer imaging modality that does not involve radiation exposure \cite{nie2016estimating}. While CT is critical for various clinical applications like radiotherapy treatment planning, there is a recent transition to MRI for these applications \cite{paczona2023magnetic}. Thus, developing algorithms that use decades of collected CT data and adapt them for newly acquired MRI scans is an important avenue of research.

The inverse problem of estimating MRI from CT is also an important application. CT is a much faster imaging modality compared to MRI, making it a better solution in emergency scenarios such as stroke. However, MRI provides more sensitive brain visualization \cite{vymazal2012comparison}. Therefore, having universal algorithms that can adapt to the needed modality at hand is highly beneficial.

While these examples highlight the clinical relevance of domain adaptation between CT and MRI, for the purpose of this benchmark, we utilize a different dataset focusing on thoracic organ segmentation. This choice is motivated by the availability of a dataset that provides both MR and CT images with corresponding segmentation maps for thoracic organs, which is essential for evaluating the performance of UDA algorithms. Despite the difference in the target application, the underlying principles of domain adaptation remain the same, and the insights gained from this benchmark can be applied to various clinical scenarios.

\paragraph{CT $\rightarrow$ low-dose CT}Second, we include a CT to low-dose CT (LDCT) shift, motivated by the increasing popularity of LDCT. LDCT produces images with a lower signal-to-noise ratio but are still diagnostically effective, resulting in several-fold lower radiation dosage exposure compared to regular CT (allowing for screening purposes \cite{lidc,kubo2016standard}), faster scanning time, mobility to scan underserved populations \cite{raghavan2020initial}, and cost-effectiveness \cite{mohammadshahi2019cost}. Similar to the CT $\leftrightarrow$ MRI domain shift, utilizing publicly available annotated regular CT scans can accelerate the development of automated segmentation models for LDCT. As demonstrated in Table \ref{tab:metrics_pure}, methods trained on regular CT perform poorly on LDCT. This shift is the only one obtained via simulation, where we algorithmically simulate low-dose CT from regular ones.

 % \reconsider{no contrast enhancement}} Why?
\paragraph{Contrast enhancement $\rightarrow$ no contrast enhancement} Third, we include two tasks, MRI and CT, involving domain transfer from a contrast-enhanced (CE) image to an image without contrast enhancement (native). CE injection is a labor-intensive step, requiring additional training for personnel and carrying a small but additional risk for patients \cite{andreucci2014side,costelloe2020risks}. Again, we suggest benchmarking DA methods against the scenario where models that utilized richer imaging modalities (CE) during supervised training are adapted for safer modalities (non-contrast-enhanced).

\paragraph{MRI settings} Finally, we include three setups that address the domain shifts caused by variable MRI scanner settings, which are among the most common sources of domain shift encountered in practice \cite{yan2020mri,medim_da_survey_2023}. These setups cover different field strengths (1.5T vs. 3T), different scanner manufacturers, and a combination of both. Domain shifts arising from variations in scanner settings are ubiquitous in multi-source MRI datasets, as differences in field strength and manufacturer-specific acquisition parameters can significantly impact the appearance and quality of the resulting images. Addressing these shifts is crucial for developing robust and generalizable segmentation models that can handle the heterogeneity of MRI data encountered in real-world clinical scenarios.

\subsection{Datasets selection}
\label{ssec:data}
% \reconsider{Intro sentence, may be add link to some medim datasets summary paper/github, comment on what is necessary to use dataset for DA setup (annotation, metadata)}
We base the inclusion of datasets into the benchmark on two criteria. \textbf{Relevance}, we aim to cover as many relevant domain shifts as possible; see Section~\ref{ssec:limitations} for a list of domain shifts not included in our benchmark. \textbf{Scale}, we prefer a dataset with a larger number of samples, when deciding between two datasets that are both relevant and include similar domain shifts. 
% \textbf{Economy}, we aim to cover selected shifts with a preferably smaller number of datasets.
All reviewed and selected datasets are summarized in Table~\ref{tab:datasets}, and all technical details (e.g., links, download instructions, and licenses) are provided in Supplementary materials.

\input{tables/3_setups}

We start by selecting a dataset for MRI to CT conversion. This allows for several alternatives. Many authors use datasets such as BTCV and CHAOS for these tasks, both of which include images of the thoracic region. BTCV consists of 30 CT scans with 13 organ annotations, and CHAOS of 40 MRI and 40 CT scans with 4 organ annotations. Another option is MM-WHS, which consists of 20 MRI and 20 CT scans of the heart with 8 annotated classes. Finally, there is the newer AMOS dataset, which consists of 500 CT and 100 MRI scans with 15 annotated thoracic organs. Following our criteria, we include AMOS as it is the largest option. We also use AMOS to simulate LDCT data.

To cover CE CT to native CT task, we add Lung Image Database Consortium image collection (LIDC) \cite{lidc}. LIDC contains chest CT images with and without contrast enhancement with segmentation annotation of lung nodules. LIDC dataset covers lung nodules - an oncology pathology, one of the most common reasons for using contrast enhancement \cite{purysko2016does}. Then, we cover the similar CE-based data shift in multi-sequence MRI data, from T1 CE to T1 modality. BraTS 2021, being one of the largest and most widely used datasets in the medical imaging community, emerges as the natural choice, satisfying our criteria of relevance, and scale.

Finally, we cover variability in the single-sequence MRI acquisition. As evident from Tables \ref{tab:benchmarks} and \ref{tab:datasets}, common choices are the ACDC and M\&Ms datasets. Both include segmentation classes of the heart and are relatively large, consisting of 150 and 375 annotated samples, respectively. Both have several concretely defined MRI domains (e.g., different scanners, parameters, field strengths). Another option is CC359, which has the same rich variability in MRI parameters and is similarly sized, including 359 annotated samples. Both ACDC and M\&Ms images have $1\times 1\times 9~ \text{mm}^3$ spacing, for which a 2D algorithm would often be a more viable choice. In contrast, a significant advantage of CC359 is the fine-grade and consistent voxel spacing, approximately $1\times 1\times 1~ \text{mm}^3$, which concludes our selection. UDA setups from the selected datasets are summarized in Table~\ref{tab:setup}.

%% file: tables/1_benchmarks.tex
\begin{table*}[ht]
    \centering
    \caption{Comparison to the existing benchmark (CrossMoDA) and datasets that are commonly used for Domain Adaptation in 3D medical image segmentation. Our proposed benchmark (M3DA) covers all primary domain shifts with the largest publicly available datasets.}

    \resizebox{\textwidth}{!}{%
    \begin{tabular}{lccccc}
        \toprule
        \textbf{Paper} & \textbf{Domain shifts} & \textbf{Datasets} & \textbf{Modalities} & \textbf{Region of interest} \\
        \midrule
        Jiang at al., 2020 \cite{jiang2020unified} & inter-modality & BTCV, CHAOS & MRI, CT &  4 thoracic organs  \\
        Al et al., 2021 \cite{al2021olva} & inter-modality & MM-WHS & MRI, CT & heart \\
        Weihsbach at al., 2024 \cite{dg_tta} & inter-modality & AMOS, MM-WHS,  spine & MRI, CT & 15 thoracic organs, brain, spine \\
        Liu et al., 2020 \cite{liu2020shape}             & MRI settings, scanners & 6 public datasets      & MRI              &  prostate         \\
        Gu et al., 2021 \cite{gu2021domain}             & MRI settings, scanners & SAML                   & MRI              &  prostate         \\
        Chen et al., 2022 \cite{chen2022maxstyle}         & MRI settings, scanners & ACDC, M\&Ms            & MRI              &  heart            \\
        Lennartz et al., 2023 \cite{lennartz2023segmentation} & MRI settings, scanners & CC359, ACDC, M\&Ms     & MRI              &  brain, heart     \\
        Wong et al., 2023 \cite{wong2023hartleymha}       & MRI settings, scanners & BraTS                  & MRI              &  brain            \\
        Liu et al., 2022 \cite{liu2022act}               & MRI inter-sequence & BraTS                  & MRI              &  brain            \\
        CrossMoDA, 2023 \cite{crossmoda}                & MRI inter-sequence & Vestibular Schwannoma  & MRI              &  brain            \\
        
        \midrule
        
        \multirow{2}{*}{M3DA (ours)} & inter-modality, inter-sequence & \multirow{2}{*}{AMOS, CC359, BraTS, LIDC} & \multirow{2}{*}{MRI, CT} & brain, tumors, \\
        & CT and MRI settings, scanners, contrast & & & 15 thoracic organs \\
        
        \bottomrule
    \end{tabular}}
    
    \label{tab:benchmarks}
\end{table*}

%% file: tables/2_datasets.tex
\begin{table}[ht]
    \centering
    \caption{Summary of the datasets commonly used in DA studies. ROI stands for region of interest, and \#cls denotes the number of foreground segmentation classes. Several datasets do not contain inner domain shifts, i.e., \textit{single source}; they are used in multi-dataset setups.}

    \resizebox{\linewidth}{!}{%
    \begin{tabular}{@{}lcccc@{}}
    \toprule
    \textbf{Dataset} & \textbf{Modality} & \textbf{ROI} & \textbf{\#cls} & \textbf{\#cases} \\
    \midrule
    WMH \cite{wmh} & MRI & Brain & 2 & 60 \\
    BraTS \cite{brats} & MRI & Brain & 3 & 1251 \\
    CC359 \cite{cc359} & MRI & Brain & 5 & 359 \\
    ACDC \cite{bernard2018deep} & MRI & Heart & 2 & 150 \\
    M\&Ms \cite{campello2021multi} & MRI & Heart & 6 & 375 \\
    SCGM \cite{prados2017spinal} & MRI & Spinal Cord & 2 & 80 \\
    IVDM3Seg \cite{IVDM3Seg2018} & MRI & Spinal Cord & 2 & 16 \\
    BTCV \cite{btcv} & CT & Abdomen & 13 & 30 \\
    AMOS \cite{amos} & CT, MRI & Abdomen & 15 & 500, 100 \\
    CHAOS \cite{kavur2021chaos} & CT, MRI & Thorax & 4 & 50 \\
    MM-WHS \cite{zhuang2016multi} & CT, MRI & Heart & 8 & 20, 20 \\
    \bottomrule
    \end{tabular}}
    \label{tab:datasets}
\end{table}

%% file: tables/3_setups.tex
\begin{table}[h]
    \centering
    \caption{Details of eight tasks in M3DA benchmark. The last three columns correspond to the numbers of cases in \textit{source}, \textit{target train}, and \textit{target test} folds, respectively.}

    \resizebox{\linewidth}{!}{%
    \begin{tabular}{@{}lccccc@{}}
        \toprule
        \textbf{Task name} & \textbf{Domain shift} & \textbf{Dataset} & $|X^s|$ & $|X^t_{tr}|$ & $|X^t_{ts}|$ \\
        \midrule
        MR$\ra$CT & inter-modality & AMOS & 60 & 200 & 150 \\
        CT$\ra$MR & inter-modality & AMOS & 150 & 40 & 60 \\
        CT$\ra$LDCT & CT settings & AMOS & 150 & 200 & 150 \\
        CE CT $\ra$ CT & settings (contrast) & LIDC & 289 & 297 & 297 \\
        % T1$\ra$T2 & intra-modality & BraTS & 625 & 313 & 313 \\
        % T2$\ra$T1 & intra-modality & BraTS & 625 & 313 & 313 \\
        % T1$\ra$T1c & settings (contrast) & BraTS & 625 & 313 & 313 \\
        T1 CE$\ra$T1 & settings (contrast) & BraTS & 625 & 313 & 313 \\
        T1 F & MRI settings & CC359 & 30 & 30 & 30 \\
        % T1 Sc & MRI scanners & CC359 & 30 & 20 & 21 \\  % add unlbl
        T1 Sc & MRI scanners & CC359 & 30 & 30 & 30 \\
        T1 Mix & settings, scanners & CC359 & 29 & 30 & 30 \\
        \bottomrule
    \end{tabular}}
    
    \label{tab:setup}
\end{table}

%% file: content/3_methods.tex
\section{Methods for M3DA Benchmark}
\label{sec:methods}

\begin{figure}[!ht]
    \centering
    \includegraphics[width=1\columnwidth]{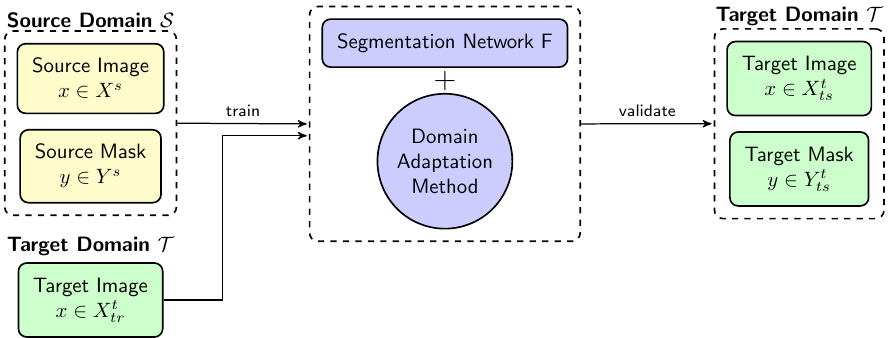}
    \caption{Overview of the UDA pipeline for semantic segmentation. Some methods does not require Target Domain images during training, e.g., nnAugm, IN, AdaBN.}
    \label{fig:pipeline}
\end{figure}

\subsection{Segmentation baseline and oracle}
\label{ssec:baseline}

% We start with a formal definition of the semantic segmentation task in M3DA.
Let $F$ be a segmentation network which takes an image $x \in \R^{H \times W \times D}$ and predicts a soft-segmentation map $p = F(x)$, $p \in \R^{H \times W \times D \times (C + 1)}$. Here, the last layer of $F$ is softmax which outputs a $(C+1)$-dimensional voxel-wise vector $\left[p^{(h, w, d, c)}\right]_c$ behaving as a discrete distribution over classes. The parameters $\theta_F$ of $F$ are learned to minimize some segmentation loss $\mathcal{L}_{seg} (p, y)$. In our case, $\mathcal{L}_{seg}$ is a sum of cross-entropy and Dice losses, as used by default in \cite{nnunet} and many other works. Optimization problem for training on source domain reads: % $\displaystyle \min_{\theta_F} \frac{1}{|X^s|} \sum_{(x, y) \in (X^s, Y^s)} \mathcal{L}_{seg} (F(x), y)$.

\begin{equation}
    \min_{\theta_F} \frac{1}{|X^s|} \sum_{(x, y) \in (X^s, Y^s)} \mathcal{L}_{seg} (F(x), y).    
\end{equation}

Further, every DA method depends on $F$, i.e., has the same backbone, so the choice of $F$ is crucial for the benchmark construction. We used the nnU-Net \cite{nnunet} architecture, loss function, and training pipeline, since nnU-Net demonstrated the best performance in several relevant tasks \cite{nnunet,amos,isensee2024nnu}, including AMOS and BraTS. We also compared nnU-Net to its closest alternatives, UNETR \cite{unetr}, Swin UNETR \cite{swinunetr}, and MedNeXt \cite{mednext}, directly within our benchmark tasks and confirmed superior nnU-Net performance; see Table~\ref{tab:backbones}.% Supplementary materials for the backbone comparison.

% \input{tables/4_backbones}

% (as DA methods)
To conduct an ablation study of normalization techniques and nnU-Net pipeline components, we replaced the default instance with batch normalization layers. We also removed modality-specific preprocessing, postprocessing, and test-time augmentations, so we could assess the unhindered impact of DA methods. A detailed technical description is given in Supplementary materials.

An nnU-Net pipeline with the changes above is a backbone for all further experiments; we call it simply U-Net. Finally, we define two core methods of the M3DA benchmark. \textbf{Baseline} -- U-Net trained on $(X^s, Y^s)$ and tested on $(X^t_{ts}, Y^t_{ts})$, i.e., naive transfer. \textbf{Oracle} -- U-Net trained and tested via cross-validation on $(X^t_{ts}, Y^t_{ts})$ that might be interpreted as an upper bound estimation for DA methods.
%, i.e., upper bound for DA methods.

Given baseline and oracle scores, the goal of DA methods therefore is to close the gap between them.% DA methods are based on the same 

\subsection{UDA methods}
\label{ssec:uda}

In our methods selection, we mainly follow reviews of DA for medical image analysis \cite{medim_da_survey_2021,medim_da_survey_2023}. We include core methods of DA for open-world images, following the corresponding reviews \cite{da_survey_2018,uda_survey_2020,uda_segm_review_2020}, top-performing solutions to the CrossMoDA challenge \cite{crossmoda}, and most recent Domain Generalization methods \cite{dg_tta}, totaling 12 methods: Adaptive Batch Normalization (AdaBN) \cite{adabn}, Instance Normalization (IN) \cite{instance_norm}, Self-ensembling (SE) \cite{se,se_medim}, Minimizing entropy (MinEnt) \cite{entropy}, Domain Adversarial Neural Network (DANN) \cite{dann,dann_medim}, CycleGAN 2D \cite{cyclegan}, CycleGAN 3D \cite{cyclegan3d}, Histogram matching (HM), nnU-Net augmentations (nnAugm) \cite{nnunet}, Gamma correction augmentation (Gamma), Global intensity non-linear (GIN) augmentation \cite{gin}, and Modality independent neighborhood descriptor (MIND) augmentation \cite{dg_tta}. Finally, we include pretrained backbones from foundational models for 3D medical imaging CLIP-driven universal model (UniModel) \cite{unimodel} and SAM-Med3D \cite{sammed}. A complete description of the methods selection is provided in Supplementary materials.

To ensure fair comparison, we maintained consistent training protocols across all methods, using U-Net backbone from nnU-Net framework for most of the experiments. Most of the DA methods require only minor architectural changes or no change at all, with only exception being DANN. Implementation details are given in Supplementary materials.

%% file: content/4_experiments.tex
\section{Experiments}
\label{sec:exp}

\subsection{Backbone selection} We evaluated four segmentation backbones (U-Net, UNETR, SwinUNETR, and MedNeXt) to determine an optimal baseline architecture for our experiments; see Table~\ref{tab:backbones}. For comparison, we included two pretrained models: SAM-Med3D and UniModel, which are based on UNETR and SwinUNETR architectures, respectively, and used officially published model weights.

\input{tables/4_backbones}

Our analysis of the foundational models (SAM-Med3D and UniModel) finetuned in supervised fashion yielded three key observations. First, both models demonstrated improved performance compared to their respective base architectures (UNETR and SwinUNETR). Second, we observed that both models slightly underperformed in 4 out of 5 MRI tasks compared to their counterparts trained from scratch, potentially due to their pretraining being predominantly conducted on CT data. Third, neither of them surpassed the performance of a standard U-Net trained from scratch.

MedNeXt performed on par with regular U-Net, however failed to converge on LIDC dataset on multiple runs attempts. % for unknown reason.
Based on these results, we selected the U-Net architecture as our primary backbone.

\subsection{Domain Adaptation methods on M3DA}

\input{tables/5_metrics_std}

\input{tables/6_ablation}

We evaluated various DA methods on the M3DA benchmark (Table~\ref{tab:metrics_pure}) using multi-class Dice score and the \textit{percentage of performance gap} closed between the Baseline and Oracle setups: $\displaystyle 100\times \frac{\text{Method}_{\text{Dice}} - \text{Baseline}_{\text{Dice}}}{\text{Oracle}_{\text{Dice}} - \text{Baseline}_{\text{Dice}}}$.

Our analysis begins with non-adapted networks (trained solely on source domain) evaluated on target domain images, represented by three baseline models: U-Net, UniModel (SwinUNETR), and SAM-Med3D (UNETR). Standard U-Net without adaptations failed completely on MR $\leftrightarrow$ CT tasks and showed poor performance on CT tasks (low-dose and CE), while maintaining moderate performance on MRI parameter shift tasks. In contrast, generalist models pretrained using contrastive and segment-anything approaches showed slightly inferior performance on MRI tasks but demonstrated remarkable results on CT-related tasks. Notably, UniModel achieved the best overall performance on the CE CT $\rightarrow$ CT task without adaptations, while SAM-Med3D exhibited strong performance on the CT $\rightarrow$ LDCT task. These results constitute, to our knowledge, the first empirical demonstration of zero-shot domain adaptation capabilities in foundational models for 3D medical imaging.

CycleGANs, despite their success in the CrossMoDA challenge, performed relatively poorly in our benchmark, particularly on CT-based tasks. We attribute this underperformance to the increased complexity of full-resolution CT images compared to brain MRI segmentation, including variations in size, localization regions, fine-grained details, and subtle stylistic differences.

Classical visual UDA methods (AdaBN, InstanceNorm, DANN, and Self-Ensembling) consistently outperformed the baseline, demonstrating their robustness across diverse domain shifts. 

Unexpectedly, generic augmentations (nnAugm) and even their subset, Gamma augmentation, outperformed more sophisticated methods on average. This finding strongly suggests the importance of incorporating generic augmentations into DA pipelines, which we explore in the following section.

Finally, recent Domain Generalization methods, GIN and MIND, achieved superior performance on MR $\leftrightarrow$ CT tasks, ranking first and second respectively, with relatively average results across other tasks. We note that these methods were originally developed and evaluated within the MR $\leftrightarrow$ CT setups, so increasing the diversity of a DA benchmark is useful for understanding the true method's capabilities.

\begin{figure}
    \centering
    \includegraphics[width=1\columnwidth]{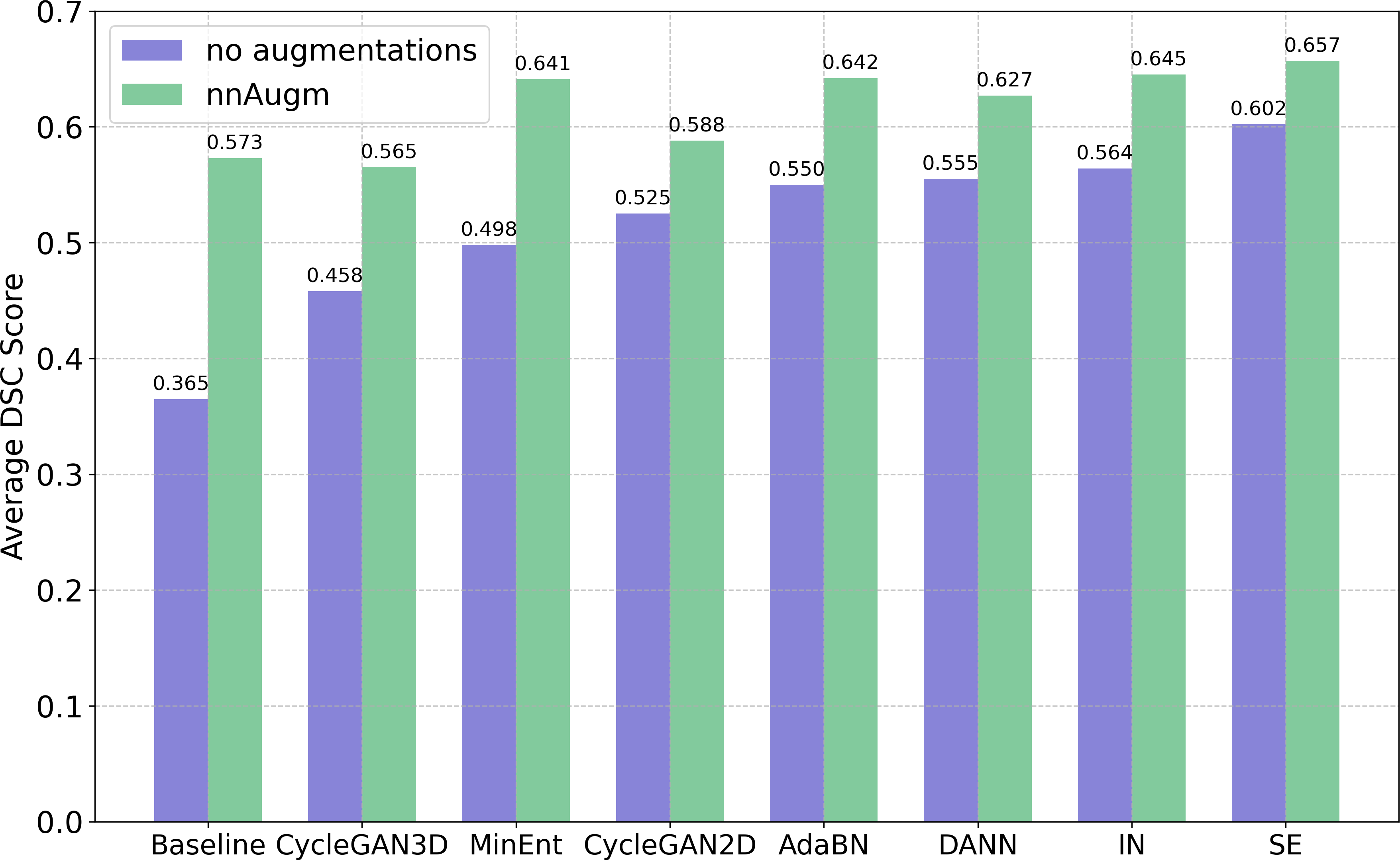}
    \caption{Comparison of DA methods with and without augmentations.}
    \label{fig:augm}
\end{figure}

%%%%%%%%%%%%%%
\subsection{Impact of additional augmentations}

Finally, we systematically evaluated the effect of incorporating nnU-Net augmentations, nnAugm, into each domain adaptation method. Our results demonstrate consistent improvements across most methods (Figure~\ref{fig:augm}) and datasets (Table~\ref{tab:ablation_aug}), with an average increase of 10.3 percentage points in Dice score. These benefits varied significantly across different domain shifts: CT-related tasks showed the most substantial improvements, while MRI-based tasks exhibited more modest gains.

Different methods also showed varying degrees of improvement when supplemented with augmentations. MinEnt demonstrated the most dramatic enhancement, +33.5 percentage points in average gap, making it the best-performing method overall (Figure~\ref{fig:teaser1}). CycleGAN-based approaches also benefited significantly, especially in CT-related tasks. Self-Ensembling, while being the strongest initial method, showed the least improvement from the additional augmentations, suggesting that it might already benefited from the incorporated augmentations by design \cite{se_medim}. Notably, some methods exhibited slight performance degradation on specific tasks, indicating that aggressive augmentation strategies may occasionally interfere with method-specific adaptation mechanisms.

Viewing these results from the alternative perspective, we can consider each method as an additional adaptation strategy applied on top of the strong nnAugm baseline. From this point of view, the marginal benefit of adding sophisticated DA methods to an already well-augmented model is more modest but still significant. The best-performing methods (SE and MinEnt) provide an additional $10$--$15\%$ improvement over nnAugm, demonstrating that DA techniques can capture domain-specific variations that generic augmentations alone cannot address. This suggests that while extensive augmentations should form the foundation of any domain adaptation pipeline, method-specific adaptation mechanisms provide complementary benefits that warrant their inclusion in the final solution.

To conclude, despite significant advances in deep learning over the past decade, our benchmark reveals a concerning trend (Figure~\ref{fig:teaser3}): DA methods for medical image segmentation have shown minimal improvement since 2017, with recent approaches performing comparably or even worse than earlier ones, suggesting that closing the domain gap remains a fundamental challenge that requires radically new approaches.

%% file: tables/4_backbones.tex
\begin{table}[h!]
    \centering
    \caption{Comparison of various backbones on M3DA benchmark in the \textit{Oracle} setup (training and validating on Target Domain). Numbers are average multiclass Dice score.}  % , so the row U-Net (ours) corresponds to the Oracle reported in the primary results.

    % \resizebox{\textwidth}{!}{%
    \resizebox{\columnwidth}{!}{%
    \begin{tabular}{lccccccccc}
        \toprule
        & \multicolumn{3}{c}{AMOS} & LIDC & BraTS & \multicolumn{3}{c}{CC359} & \\

        \cmidrule(lr){2-4} \cmidrule(lr){5-5} \cmidrule(lr){6-6} \cmidrule(lr){7-9}
        
        & CT & MR & LDCT & CT & T1 & T1 F & T1 Sc & T1 Mix & \textit{avgDSC} \\
        
        \midrule
        
        % UNETR \cite{unetr} & 0.675 & 0.791 & 0.658 & 0.240 & 0.559 & 0.953 & 0.954 & 0.957 & 0.723 \\
        UNETR & 0.675 & 0.791 & 0.658 & 0.366 & 0.559 & 0.953 & 0.954 & 0.957 & 0.738 \\
        SAM-Med3D & 0.742 & 0.813 & 0.681 & 0.437 & 0.473 & 0.949 & 0.914 & 0.951 & 0.745 \\
       
        % MedNeXt \cite{mednext} & 0.869 & 0.818 & 0.826 & 0.003 & 0.741 & 0.961 & 0.964 & 0.962 & 0.768 \\
        % MedNeXt \cite{mednext} & 0.869 & 0.818 & 0.826 & 0.016 & 0.741 & 0.961 & 0.964 & 0.962 & 0.770 \\
        MedNeXt & 0.869 & 0.818 & 0.826 & 0.000 & 0.741 & 0.961 & 0.964 & 0.962 & 0.768 \\
        % \footnote{Experiments with MedNeXt on LIDC failed to converge on several running attempts.}
        
        SwinUNETR & 0.780 & 0.791 & 0.741 & 0.448 & 0.660 & 0.954 & 0.953 & 0.957 & 0.785 \\

        UniModel & 0.819 & 0.812 & 0.776 & 0.504 & 0.630 & 0.955 & 0.920 & 0.958 & 0.797 \\
        
        U-Net & 0.842 & 0.826 & 0.814 & 0.519 & 0.686 & 0.954 & 0.957 & 0.958 & 0.820 \\

        nnU-Net & 0.879 & 0.818 & 0.848 & 0.455 & 0.739 & 0.963 & 0.963 & 0.965 & 0.829 \\
        
        \bottomrule
         
    \end{tabular}}
    \label{tab:backbones}
\end{table}

%% file: tables/5_metrics_std.tex
% \citeyear{gin}, 
\begin{table*}
    \centering
    \caption{Main results on our benchmark in terms of multiclass average Dice score, where background label is excluded from quantification. The best results in each column are highlighted in \textbf{bold}. Case-wise standard deviations for these experiments are provided in parentheses. Foundational models UniModel and SAM-Med3D were finetuned in the baseline setting, similar to U-Net.}

    \resizebox{\textwidth}{!}{%
    \begin{tabular}{lcccccccccc}
        \toprule
        & MR$\rightarrow$CT & CT$\rightarrow$MR & CT$\rightarrow$LDCT & CE CT$\rightarrow$CT & T1 CE$\rightarrow$T1 & T1 F & T1 Sc & T1 Mix & \textit{avg DSC} & \textit{avg gap} \\
        
        \midrule
        
        U-Net (Baseline)      & 0.032 (0.045) & 0.032 (0.038) & 0.133 (0.162) & 0.228 (0.265) & 0.426 (0.172) & 0.741 (0.067) & 0.766 (0.025) & 0.560 (0.159) & 0.365 & 0.0\% \\

        SAM-Med3D \cite{sammed} & 0.019 (0.013) & 0.037 (0.031) & 0.524 (0.120) & 0.412 (0.307) & 0.270 (0.155) & 0.645 (0.080) & 0.758 (0.035) & 0.486 (0.274) & 0.394 & -1.0\% \\

        UniModel \cite{unimodel} & 0.027 (0.017) & 0.012 (0.013) & 0.252 (0.191) & \textbf{0.470 (0.313)} & 0.331 (0.143) & 0.740 (0.064) & 0.736 (0.038) & 0.618 (0.179) & 0.398 & 7.4\%\\
        
        \midrule
        
        HM & 0.331 (0.182) & 0.222 (0.128) & 0.111 (0.174) & 0.133 (0.221) & 0.341 (0.183) & 0.789 (0.069) & 0.748 (0.090) & 0.504 (0.195) & 0.397 & -1.1\% \\
        
        CycleGAN 3D & 0.333 (0.128) & 0.264 (0.113) & 0.326 (0.175) & 0.130 (0.203) & 0.345 (0.175) & 0.791 (0.035) & 0.713 (0.023) & 0.762 (0.017) & 0.458 & 9.5\% \\
        
        MinEnt \cite{entropy} & 0.140 (0.136) & 0.172 (0.149) & 0.505 (0.127) & 0.392 (0.323) & 0.429 (0.168) & 0.770 (0.038) & 0.798 (0.025) & 0.776 (0.088) & 0.498 & 28.5\% \\
        
        CycleGAN 2D & 0.205 (0.153) & 0.406 (0.144) & 0.530 (0.187) & 0.216 (0.260) & 0.398 (0.181) & 0.852 (0.015) & 0.801 (0.027) & 0.795 (0.024) & 0.525 & 30.2\% \\
        
        GIN \cite{gin} & \textbf{0.589 (0.144)} & \textbf{0.637 (0.105)} & 0.722 (0.108) & 0.163 (0.238) & 0.382 (0.181) & 0.837 (0.066) & 0.709 (0.069) & 0.804 (0.062) & 0.605 & 33.6\% \\
        
        AdaBN \cite{adabn} & 0.322 (0.157) & 0.353 (0.177) & 0.587 (0.202) & 0.295 (0.291) & 0.433 (0.165) & 0.778 (0.042) & 0.833 (0.020) & 0.796 (0.059) & 0.550 & 35.0\% \\
        
        DANN \cite{dann_medim} & 0.296 (0.147) & 0.278 (0.135) & 0.699 (0.148) & 0.409 (0.297) & 0.416 (0.161) & 0.730 (0.078) & 0.833 (0.029) & 0.776 (0.082) & 0.555 & 36.2\% \\
        
        IN \cite{instance_norm} & 0.303 (0.149) & 0.308 (0.143) & 0.668 (0.167) & 0.427 (0.287) & 0.428 (0.155) & 0.756 (0.058) & 0.838 (0.025) & 0.784 (0.078) & 0.564 & 39.6\% \\
        
        MIND \cite{dg_tta} & 0.560 (0.171) & 0.588 (0.125) & 0.237 (0.148) & 0.425 (0.236) & 0.335 (0.162) & 0.865 (0.035) & 0.869 (0.039) & 0.845 (0.033) & 0.590 & 45.9\% \\
        
        Gamma          & 0.349 (0.182) & 0.166 (0.146) & 0.241 (0.230) & 0.441 (0.313) & 0.443 (0.163) & 0.893 (0.018) & \textbf{0.910 (0.006)} & 0.910 (0.012) & 0.544 & 48.3\% \\
        
        SE \cite{se_medim} & 0.391 (0.133) & 0.388 (0.101) & 0.603 (0.189) & 0.332 (0.291) & 0.388 (0.175) & 0.906 (0.023) & 0.893 (0.013) & \textbf{0.918 (0.018)} & 0.602 & 51.7\% \\
        
        nnAugm \cite{nnunet} & 0.166 (0.125) & 0.102 (0.090) & \textbf{0.779 (0.103)} & 0.392 (0.315) & \textbf{0.446 (0.164)} & \textbf{0.910 (0.010)} & 0.897 (0.012) & 0.889 (0.012) & 0.573 & 51.9\% \\
        
        % nnUNet          & 0.397 & 0.355 & 0.750 & 0.373 & 0.330 & 0.923 & 0.914 & 0.907 & 0.619 & 54.9\% \\
        
        % \midrule

        % best in setup & 0.589 & 0.637 & 0.779 & 0.470 & 0.446 & 0.910 & 0.910 & 0.918 & 0.704 & 72.9\% \\
        
        \midrule
        
        U-Net (Oracle)        & 0.842 (0.092) & 0.826 (0.035) & 0.814 (0.095) & 0.519 (0.297) & 0.686 (0.178) & 0.954 (0.017) & 0.957 (0.012) & 0.958 (0.009) & 0.820 & 100.0\% \\
        
        \bottomrule
         
    \end{tabular}}
    \label{tab:metrics_pure}
\end{table*}

%% file: tables/6_ablation.tex
\begin{table*}[ht]
    \centering
    \caption{Performance comparison of different methods supplemented with nnU-Net augmentations. Colored numbers show an improvement (or a decline, respectively) over a non-augmented method. GIN and MIND were only trained with nnU-Net augmentations.}
    
    % Add these color definitions to your preamble
    \definecolor{darkGreen}{RGB}{0, 102, 0}     % For improvements > 0.15
    \definecolor{medGreen}{RGB}{0, 153, 0}      % For improvements 0.05 to 0.15
    \definecolor{lightGreen}{RGB}{144, 238, 144} % For small improvements 0 to 0.05
    \definecolor{lightRed}{RGB}{255, 200, 200}   % For negative values
    
    \resizebox{\textwidth}{!}{%
    \begin{tabular}{lcccccccccc}
        \toprule
        & MR$\rightarrow$CT & CT$\rightarrow$MR & CT$\rightarrow$LDCT & CE CT$\rightarrow$CT & T1 CE$\rightarrow$T1 & T1 F & T1 Sc & T1 Mix & \textbf{avg DSC} & \textbf{avg gap} \\
        
        \midrule
        
        % GIN & 0.589 & 0.637 & 0.722 & 0.163 & 0.382 & 0.837 & 0.709 & 0.804 & 0.605 & 33.6\% \\
        
        CycleGAN 3D & 0.364 \textcolor{lightGreen}{$\uparrow$0.031} & 0.464 \textcolor{darkGreen}{$\uparrow$0.200} & 0.679 \textcolor{darkGreen}{$\uparrow$0.353} & 0.221 \textcolor{medGreen}{$\uparrow$0.091} & 0.379 \textcolor{lightGreen}{$\uparrow$0.034} & 0.825 \textcolor{lightGreen}{$\uparrow$0.034} & 0.810 \textcolor{medGreen}{$\uparrow$0.097} & 0.779 \textcolor{lightGreen}{$\uparrow$0.017} & 0.565 \textcolor{medGreen}{$\uparrow$0.107} & 34.1\% \textcolor{darkGreen}{$\uparrow$24.6\%} \\
        CycleGAN 2D & 0.301 \textcolor{medGreen}{$\uparrow$0.096} & 0.461 \textcolor{medGreen}{$\uparrow$0.055} & 0.666 \textcolor{medGreen}{$\uparrow$0.136} & 0.333 \textcolor{medGreen}{$\uparrow$0.117} & 0.416 \textcolor{lightGreen}{$\uparrow$0.018} & 0.865 \textcolor{lightGreen}{$\uparrow$0.013} & 0.850 \textcolor{lightGreen}{$\uparrow$0.049} & 0.815 \textcolor{lightGreen}{$\uparrow$0.020} & 0.588 \textcolor{medGreen}{$\uparrow$0.063} & 45.5\% \textcolor{medGreen}{$\uparrow$15.3\%} \\
        
        % MIND & 0.560 & 0.588 & 0.237 & 0.425 & 0.335 & 0.865 & 0.869 & 0.845 & 0.590 & 45.9\% \\
        
        Baseline (nnAugm) & 0.166 \textcolor{medGreen}{$\uparrow$0.134} & 0.102 \textcolor{medGreen}{$\uparrow$0.070} & 0.779 \textcolor{darkGreen}{$\uparrow$0.646} & 0.392 \textcolor{darkGreen}{$\uparrow$0.164} & 0.446 \textcolor{lightGreen}{$\uparrow$0.020} & 0.910 \textcolor{darkGreen}{$\uparrow$0.169} & 0.897 \textcolor{medGreen}{$\uparrow$0.131} & 0.889 \textcolor{darkGreen}{$\uparrow$0.329} & 0.573 \textcolor{darkGreen}{$\uparrow$0.208} & 51.9\% \textcolor{darkGreen}{$\uparrow$51.9\%} \\  % nnAugm (Baseline)
        DANN        & 0.414 \textcolor{medGreen}{$\uparrow$0.118} & 0.349 \textcolor{medGreen}{$\uparrow$0.071} & 0.809 \textcolor{medGreen}{$\uparrow$0.110} & 0.411 \textcolor{lightGreen}{$\uparrow$0.002} & 0.403 \textcolor{lightRed}{$\downarrow$-0.013} & 0.899 \textcolor{darkGreen}{$\uparrow$0.169} & 0.848 \textcolor{lightGreen}{$\uparrow$0.015} & 0.885 \textcolor{medGreen}{$\uparrow$0.109} & 0.627 \textcolor{medGreen}{$\uparrow$0.072} & 54.9\% \textcolor{darkGreen}{$\uparrow$23.3\%} \\

        IN          & 0.422 \textcolor{medGreen}{$\uparrow$0.119} & 0.471 \textcolor{darkGreen}{$\uparrow$0.163} & 0.796 \textcolor{medGreen}{$\uparrow$0.128} & 0.410 \textcolor{lightRed}{$\downarrow$-0.017} & 0.416 \textcolor{lightRed}{$\downarrow$-0.012} & 0.907 \textcolor{darkGreen}{$\uparrow$0.151} & 0.854 \textcolor{lightGreen}{$\uparrow$0.016} & 0.883 \textcolor{medGreen}{$\uparrow$0.099} & 0.645 \textcolor{medGreen}{$\uparrow$0.081} & 58.1\% \textcolor{darkGreen}{$\uparrow$26.6\%} \\
        AdaBN       & 0.495 \textcolor{darkGreen}{$\uparrow$0.173} & 0.532 \textcolor{darkGreen}{$\uparrow$0.179} & 0.604 \textcolor{lightGreen}{$\uparrow$0.017} & 0.365 \textcolor{medGreen}{$\uparrow$0.070} & 0.454 \textcolor{lightGreen}{$\uparrow$0.021} & 0.907 \textcolor{medGreen}{$\uparrow$0.129} & 0.890 \textcolor{medGreen}{$\uparrow$0.057} & 0.892 \textcolor{medGreen}{$\uparrow$0.096} & 0.642 \textcolor{medGreen}{$\uparrow$0.092} & 59.2\% \textcolor{darkGreen}{$\uparrow$24.2\%} \\

        SE          & 0.459 \textcolor{medGreen}{$\uparrow$0.068} & 0.571 \textcolor{darkGreen}{$\uparrow$0.183} & 0.768 \textcolor{darkGreen}{$\uparrow$0.165} & 0.389 \textcolor{medGreen}{$\uparrow$0.057} & 0.374 \textcolor{lightRed}{$\downarrow$-0.014} & 0.902 \textcolor{lightRed}{$\downarrow$-0.004} & 0.907 \textcolor{lightGreen}{$\uparrow$0.014} & 0.888 \textcolor{lightRed}{$\downarrow$-0.030} & 0.657 \textcolor{medGreen}{$\uparrow$0.055} & 60.1\% \enspace \textcolor{medGreen}{$\uparrow$8.4\%} \\
        MinEnt      & 0.388 \textcolor{darkGreen}{$\uparrow$0.248} & 0.362 \textcolor{darkGreen}{$\uparrow$0.190} & 0.788 \textcolor{darkGreen}{$\uparrow$0.283} & 0.449 \textcolor{medGreen}{$\uparrow$0.057} & 0.448 \textcolor{lightGreen}{$\uparrow$0.019} & 0.903 \textcolor{medGreen}{$\uparrow$0.133} & 0.901 \textcolor{medGreen}{$\uparrow$0.103} & 0.892 \textcolor{medGreen}{$\uparrow$0.116} & 0.641 \textcolor{medGreen}{$\uparrow$0.143} & 62.0\% \textcolor{darkGreen}{$\uparrow$33.5\%} \\

        \midrule
        % \textbf{Average improvement} & 0.376 \textcolor{medGreen}{$\uparrow$0.123} & 0.414 \textcolor{darkGreen}{$\uparrow$0.139} & 0.736 \textcolor{darkGreen}{$\uparrow$0.230} & 0.371 \textcolor{medGreen}{$\uparrow$0.068} & 0.417 \textcolor{lightGreen}{$\uparrow$0.009} & 0.890 \textcolor{medGreen}{$\uparrow$0.099} & 0.870 \textcolor{medGreen}{$\uparrow$0.060} & 0.865 \textcolor{medGreen}{$\uparrow$0.095} & 0.617 \textcolor{medGreen}{$\uparrow$0.103} \\
        \textbf{average} &\textcolor{medGreen}{$\uparrow$0.123} & \textcolor{darkGreen}{$\uparrow$0.139} & \textcolor{darkGreen}{$\uparrow$0.230} & \textcolor{medGreen}{$\uparrow$0.068} & \textcolor{lightGreen}{$\uparrow$0.009} & \textcolor{medGreen}{$\uparrow$0.099} & \textcolor{medGreen}{$\uparrow$0.060} & \textcolor{medGreen}{$\uparrow$0.095} & \textcolor{medGreen}{$\uparrow$0.103} & \textcolor{darkGreen}{$\uparrow$26.0\%} \\  % 53.2\%
        
        \bottomrule
    \end{tabular}}
    \label{tab:ablation_aug}
\end{table*}

%% file: content/5_discussion.tex
\section{Discussion}
\label{sec:discussion}

While we focused our computational experiments on unsupervised DA, M3DA also supports other DA frameworks.

\paragraph{Supervised DA} involves having annotated data from source and target domains. It can potentially close the performance gap more effectively, leveraging the explicit knowledge of target domain characteristics. All datasets and samples in M3DA come with segmentation annotations, allowing supervised DA setup.

\paragraph{Source-free DA} In this setting, the model is trained on the source domain and later adapted to the target domain without accessing source data. This approach is particularly relevant in scenarios with privacy concerns. M3DA allows source-free DA by removing the source data during finetuning.

\paragraph{Test-time DA} focuses on adapting the model during inference. This method adjusts to the target domain using only the data available at the inference time. Similar to source-free DA, one can limit access to the source domain and use online sampling of the target data.

\paragraph{Domain Generalization (DG)} aims to learn domain-invariant features from multiple source domains without accessing any target domain data during training. This approach is particularly valuable in medical imaging where encountering completely new domains is common, such as images from different hospitals or scanner manufacturers. M3DA's diverse collection of datasets from various medical centers and imaging protocols\footnote{LIDC is sourced from seven academic centers and eight medical imaging companies, BraTS is sourced from at least nine different clinical centers, AMOS was collected in two medical centers, from eight different scanners} makes it well-suited for developing and evaluating DG methods.

These alternative DA frameworks showcase the versatility of our benchmark and its potential to support a wide range of research questions and methodologies in the field of domain adaptation for medical image segmentation.

\subsection{Limitations and future directions}
\label{ssec:limitations}

While we incorporate a diverse set of domain shifts, our benchmark is not exhaustive. We exclude several candidate domain shifts due to their simplicity, low relevance, or lack of available public data. First, the CT reconstruction kernel (from sinogram space to voxel space) is an important parameter. As shown in \cite{fbpaug}, this problem can be largely mitigated via simple augmentations or an auxiliary loss function \cite{shimovolos2022adaptation}. We include our results on this shift in Supplementary materials, but as it is almost fully addressed by simple augmentations, we exclude it from the main benchmark. Also, public datasets containing both reconstruction kernel information and segmentation annotations are scarce.

Second, the CC359 dataset consists of data from six distinct domains, allowing for 30 different domain shift scenarios. We selected the three least ``solved'' shifts based on our preliminary analysis. A complete table with results on all 30 domain pairs is provided in Supplementary materials.

Third, a common critique of the BraTS dataset is that it is heavily preprocessed. Incorporating raw datasets like Burdenko-GP \cite{Zolotova2023Burdenko} could provide an evaluation of DA methods in a more realistic setting. However, this comes with an inevitable trade-off of added complexity in data preparation. We also exclude the MRI T1 to T2 domain shift from the main paper. Although we provide the results for this setup in Supplementary materials, we did not find sufficient evidence to support its clinical relevance, leading to its exclusion from our benchmark.

Finally, while the LIDC dataset allows for the CE CT to CT shift, it is primarily designed for object detection tasks, with multiple nodules per image. Despite this limitation, LIDC remains the only publicly available dataset of sufficient size that enables this clinically relevant domain shift.

Future work should focus on expanding M3DA to include more clinically relevant domain shifts, incorporating datasets with unprocessed scans, and exploring novel approaches to domain adaptation.

\subsection{Conclusion}
\label{sec:conclusion}

In this paper, we introduced the M3DA benchmark for unsupervised domain adaptation in 3D medical image segmentation. Addressing the widely indicated need for developing DA methods in the medical imaging domain \cite{gulrajani2020search,uda_survey_2020,zhuang2020comprehensive,peng2018visda,zhang2021empirical}, we created a large-scale benchmark to facilitate the development of robust segmentation methods in such a crucial application area. Contrary to previously used setups, we covered a diverse set of domain shift sources while using large, publicly available datasets.

We benchmarked the core adaptation methods, covering all key categories of UDA approaches \cite{medim_da_survey_2021,medim_da_survey_2023}, and medical foundational models. Our results revealed that adapted segmentation models struggle to generalize beyond their training distribution when tested at scale. Although some DA methods showed promise in particular settings, we revealed they might completely fail in a number of other setups. This highlights a pressing need for creating robust methods for medical image segmentation, and we hope to foster research efforts in improving adaptation methods’ performance in diverse situations.

Finally, we described several alternative problem settings within M3DA, e.g., supervised and test-time DA, enabling the evaluation of more complex hypotheses across a wider spectrum of DA methods.

%% file: content/X_suppl.tex
\clearpage
\maketitlesupplementary

\input{content/suppl/1_datasets}

\input{content/suppl/2_methods}

\input{content/suppl/3_experiments}

%% file: content/suppl/1_datasets.tex
\begin{figure*}[ht]
    \centering
    \includegraphics[width=1\linewidth]{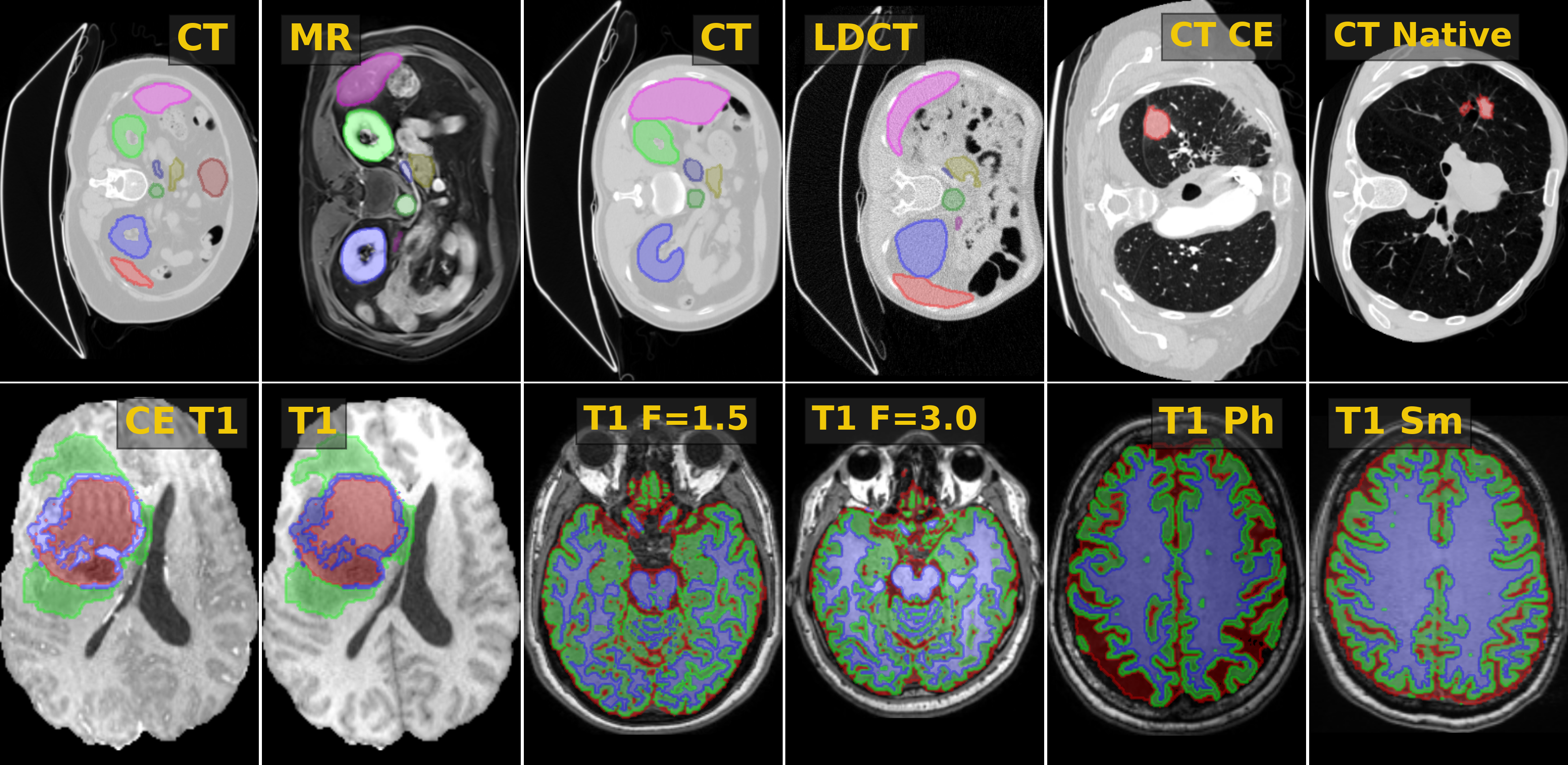}
    \caption{Examples from individual domains in M3DA with the corresponding segmentation masks. Left to right, top to bottom: CT to MR, CT to LDCT, CT CE to CT native, CE T1 to T1, T1 Field (1.5T to 3T), T1 Scanner (Philips to Siemens). Different colors correspond to different segmentation classes.}
    \label{fig:contours}
\end{figure*}

\section{Datasets description}

Below we provide an extended description of datasets used in M3DA benchmark, download and usage examples are available at \href{https://github.com/BorisShirokikh/M3DA}{https://github.com/BorisShirokikh/M3DA}. Example 2D slices from every dataset for visual comparison between domains are given in Figure~\ref{fig:contours}. Summary of licenses and data access is given in Table~\ref{tab:supp_datasets}.

\input{tables/suppl/1_datasets}

\subsection{AMOS}

The AMOS dataset \cite{amos} contains 500 CT and 100 MRI abdominal scans with the multi-organ segmentation task: liver, stomach, spleen, left and right kidneys, bladder, aorta, pancreas, inferior vena cava, duodenum, prostate/uterus, gallbladder, esophagus, left and right adrenals . As a largest available dataset for inter-modality segmentation, we employed it in MR$\ra$CT and CT$\ra$MR domain shift setups.

Furthermore, we used AMOS CT images to create one of the most clinically relevant domain shift setups -- difference in the radiation dose during scanning. 
 
For the LDCT domain, we simulated low radiation dose using the algorithm provided in \cite{ldct}, simulated data are available at \href{https://zenodo.org/records/13373720}{https://zenodo.org/records/13373720}.

\subsection{BraTS}

BraTS \cite{brats} is comprised of 2000 brain MRI cases, each consisting of four sequences: T1, T1c, T2, FLAIR, with a glioblastoma segmentation classes (3 foreground classes and background).  We only used 1251 cases with publicly available annotations and T1, T1c MRI sequences for T1 CE$\ra$T1 shift. Since sequences of the same case provide information about the same subject, we ensured source-target splits so that every case falls into exactly one fold.

\subsection{CC359}

The CC359 dataset \cite{cc359} contains 359 brain MR T1 images from three scanners, namely, GE, Philips (PH), and Siemens (SM), obtained using two magnetic field strength values, $1.5$ and $3.0$T. The dataset can be split into six domains defined by two different field strengths $\times$ three vendors, each with approximately 60 images, so it yields 30 possible domain adaptation pairs. 

CC359 also offers three tasks: brain, hippocampus, and white matter, gray matter, and cerebrospinal fluid (WMGMCSF) segmentation. We ommited hippocampus segmentation task from the benchmark, because our preliminary experiments showed it is not significantly affected by domain shifts, the relative performance drop is less than $2\%$ in every domain pair; see Table~\ref{tab:hippo}. We also omitted the brain segmentation task for the same reason, see results in \cite{shirokikh2020first}.

\input{tables/suppl/2_cc359_hippo}

Therefore, we focus only on the WMGMCSF segmentation task in CC359: white matter, gray matter and cerebral spinal fluid segmentation classes and background. From 30 possible domain pairs, we selected three with the maximum performance drop, highlighted in \textbf{bold},  Table~\ref{tab:wmgmcsf}): changing field strength with a fixed scanner PH 1.5T $\ra$ PH 3.0T (drop from 95.4 to 74.1 Dice score), changing scanner with the fixed field strength PH 3.0T $\ra$ SM 3.0T (drop from 95.7 to 76.6), and changing both parameters SM 3.0T $\ra$ GE 1.5T (drop from 95.8 to 56.0). We denote them as T1 F, T1 Sc, and T1 Mix, respectively.

\input{tables/suppl/3_cc359_wmgmcsf}

\subsection{LIDC}

LIDC \cite{lidc} is a multi-center collection of diagnostic and lung cancer screening thoracic CT scans with annotated lesions. It includes 1308 studies (of which 1018 include CT studies) from 1010 patients. Lung's nodules is one of the few clinical applications where both CE CT and CT are used, first for the initial scan, and second for the follow-ups \cite{purysko2016does}. We used LIDC for CE CT $\ra$ CT domain shift, we split data into three roughly equal groups, ommiting scans with empty masks: contrast enhanced CT (source domain) $X^s$, CT without contrast enhancement $X^t_{tr}$ (training part, target domain), and CT without contrast enhancement  $X^t_{ts}$ (test part, target domain). $X^t_{tr}$ and $X^t_{ts}$ were stratified by the number of lesions.

%% file: tables/suppl/1_datasets.tex
\begin{table}[h]
    \centering
    \caption{Datasets licenses and independent source links.}

    \resizebox{\linewidth}{!}{%
    \begin{tabular}{@{}lcc@{}}
    \toprule
    \textbf{Dataset} & license & link to dataset \\
    \midrule
    BraTS \cite{brats} & CC BY 4.0  & \href{https://www.cancerimagingarchive.net/analysis-result/rsna-asnr-miccai-brats-2021/}{https://www.cancerimagingarchive.net/analysis-result/rsna-asnr-miccai-brats-2021/} \\
    
    CC359 \cite{cc359} & CC BY-ND 4.0 & \href{https://www.ccdataset.com/download}{https://www.ccdataset.com/download}  \\
    
    AMOS \cite{amos} & CC BY 4.0 & \href{https://zenodo.org/records/7262581}{https://zenodo.org/records/7262581} \\

    AMOS LDCT & CC BY 4.0 & \href{https://zenodo.org/records/13373720}{https://zenodo.org/records/13373720} \\
    
    LIDC \cite{lidc} & CC BY 3.0 & \href{https://www.cancerimagingarchive.net/collection/lidc-idri/}{https://www.cancerimagingarchive.net/collection/lidc-idri/}\\
    
    \bottomrule
    \end{tabular}
    }
    \label{tab:supp_datasets}
\end{table}

%% file: tables/suppl/2_cc359_hippo.tex
\begin{table}[!htb]
    \centering
    \caption{Baseline and oracle results on the CC359 hippocampus segmentation task.}
    \label{tab:hippo}
    \resizebox{\columnwidth}{!}{%
    \begin{tabular}{|c|l||c|c|c|c|c|c|}

        \hline
        \multicolumn{2}{|l||}{\multirow{2}{*}{}}  & \multicolumn{6}{c|}{Target domains}\\ 
        \cline{3-8}
        \multicolumn{2}{|l||}{} & GE1.5 & PH1.5 & SM1.5 & GE3.0 & PH3.0 & SM3.0 \\ 
        \hline
        \hline

        % \clrtb

        % \multirow{6}{*}{{\STAB{\rotatebox[origin=c]{90}{Source domains}}}}
        \multirow{6}{*}{{\rotatebox[origin=c]{90}{Source domains}}}
        
        & GE 1.5 & 92.3 & 86.7 & 88.7 & 87.8 & 91.2 & 91.2 \\
        \cline{2-8}
        
        & PH 1.5 & 91.3 & 86.9 & 87.7 & 87.7 & 89.7 & 89.9 \\
        \cline{2-8}
        
        & SM 1.5 & 91.7 & 86.6 & 89.3 & 88.2 & 90.9 & 90.8 \\
        \cline{2-8}
        
        & GE 3.0 & 91.4 & 86.4 & 88.0 & 89.1 & 90.5 & 91.3 \\
        \cline{2-8}
        
        & PH 3.0 & 91.5 & 86.5 & 88.3 & 87.7 & 92.0 & 91.0 \\
        \cline{2-8}
        
        & SM 3.0 & 90.8 & 86.5 & 87.8 & 88.0 & 90.6 & 92.1 \\
        \hline
        
    \end{tabular}}
\end{table}

%% file: tables/suppl/3_cc359_wmgmcsf.tex
\begin{table}[!htb]
    \centering
    \caption{Baseline and oracle results on the CC359 WMGMSCF segmentation task.}\label{tab:wmgmcsf}
    \resizebox{\columnwidth}{!}{%
    \begin{tabular}{|c|l||c|c|c|c|c|c|}

        \hline
        \multicolumn{2}{|l||}{\multirow{2}{*}{}}  & \multicolumn{6}{c|}{Target domains}\\ 
        \cline{3-8}
        \multicolumn{2}{|l||}{} & GE1.5 & PH1.5 & SM1.5 & GE3.0 & PH3.0 & SM3.0 \\ 
        \hline
        \hline

        % \clrtb

        % \multirow{6}{*}{{\STAB{\rotatebox[origin=c]{90}{Source domains}}}}
        \multirow{6}{*}{{\rotatebox[origin=c]{90}{Source domains}}}
        & GE 1.5 & 95.8 & 82.1 & 90.8 & 82.1 & 92.6 & 80.8 \\
        \cline{2-8}
        
        & PH 1.5 & 80.1 & 92.7 & 90.8 & 93.4 & \textbf{74.1} & 90.1 \\
        \cline{2-8}
        
        & SM 1.5 & 89.7 & 85.3 & 95.6 & 86.2 & 86.2 & 84.5 \\
        \cline{2-8}
        
        & GE 3.0 & 76.6 & 89.9 & 90.3 & 95.9 & 72.0 & 91.4 \\
        \cline{2-8}
        
        & PH 3.0 & 90.6 & 74.7 & 86.0 & 75.4 & 95.4 & \textbf{76.6} \\
        \cline{2-8}
        
        & SM 3.0 & \textbf{56.0} & 88.6 & 84.9 & 92.4 & 68.4 & 95.7 \\
        \hline
        
    \end{tabular}}
\end{table}

%% file: content/suppl/2_methods.tex
\section{Methods details}

\subsection{UDA methods selection}

\paragraph{Discrepancy-based approaches} are based on incorporating maximum mean discrepancy measure as a regularization or auxilary loss function \cite{mmd_ghifary2014domain,mmd_tzeng2014deep,mmd_long2015learning}. These approaches were soon surpassed by simpler approximations, such as \text{DeepCORAL}~\cite{deepcoral}. However, all of them become computationally intractable due to significantly larger feature space in 3D segmentation task.% We thus selected more practical approaches to discrepancy minimization.
% (dim 1e6 instead 1e3 for feature space)

Since batch normalization (BN) \cite{bn} became the standard in DL, it allowed to reduce covariate shift by aligning first and second moments of feature distributions. But it introduced discrepancy between train and test by applying train-estimated statistics to the test samples. Here, \textit{Adaptive BN (AdaBN)} \cite{adabn} recalculates BN statistics on the unlabeled target data, helping to adapt to the target domain.

Contrary, \textit{Instance Normalization (IN)} \cite{instance_norm} was proposed for an efficient image stylization, and it calculates statistics for every input independently. This way IN might help adaptation, so we included IN to test it separately.

\textbf{Selected methods:} AdaBN, IN.

\paragraph{Self-training} uses predicted pseudo-labels on the target data to regularize the downstream model. For instance, the authors of \cite{se} proposed \textit{self-ensembling (SE)} for visual DA. The same methodology was implemented for 3D medical image segmentation in \cite{se_medim}. The authors trained the first, student, network on the downstream task and updated the weights of the second, teacher, network via exponential moving average. They additionally imposed a consistency criterion: mean squared error between predictions of the two networks, thus, student network minimizes segmentation and consistency losses. We included SE with hyperparameters recommended in \cite{se_medim}.

Specifically for semantic segmentation, training on self-generated predictions was shown to help in DA \cite{self_training}. Later, the authors of \cite{entropy} noted the connection between self-training and entropy minimization. They also showed that \textit{minimizing the entropy (MinEnt)} of predictions surpasses self-training and other DA methods, so we included MinEnt in our benchmark.

\textbf{Selected methods:} SE, MinEnt.

% (or maximizing the loss)
\paragraph{Adversarial-based approaches} form the basis for the most DA methods, as shown in \cite{uda_segm_review_2020}. The central idea is reversing the gradient from the domain classification network, thus learning domain invariant features for source and target inputs. To this end, the authors of \cite{dann} proposed \textit{Domain Adversarial Neural Network (DANN)} for image classification, noting that their approach is generic and can handle any output label space. Consequently, DANN was implemented for DA in 3D medical image segmentation \cite{dann_medim}.

Although many other DANN modifications exist, e.g., decoupling feature encoders for source and target images \cite{adda} or connecting the domain classification network to the output layer \cite{tsai2018learning}, adapting them for 3D segmentation requires a separate effort. Hence, we focused on testing the core method and proceeded with the close to original DANN implementation of \cite{dann_medim}.

\textbf{Selected methods:} DANN.

\paragraph{Image-level adaptation} is typically achieved using Generative Adversarial Network (GAN) \cite{goodfellow2020generative}. The goal is to learn a mapping function between the source and target domains with a generator network. Then, one can use this generator to transfer images styles between domains. Specifically for UDA, the authors of \cite{cyclegan} proposed \textit{CycleGAN 2D} which additionally enforces the reconstruction loop consistency upon two generators. This method was also designed for 3D medical images in \cite{cyclegan3d}; and it found numerous successful applications to medical image segmentation, e.g., top-3 solutions of the CrossMoDA challenge \cite{crossmoda} used \textit{CycleGAN 3D}. We included both approaches.

Image-level adaptation also includes non-generative approaches, such as Fourier Domain Adaptation (FDA) \cite{fda}, where the style of images is changed by substituting their low frequencies in Fourier space. The authors of \cite{fda_medim} succeeded in applying FDA to 3D medical image segmentation. However, such methods, similar to CT reconstruction kernel modulation \cite{fbpaug}, are not generic and heavily depend on modality-specific features, so we excluded them from further consideration.

\textbf{Selected methods:} CycleGAN 2D, CycleGAN 3D.

\paragraph{Preprocessing and augmentation} are often overlooked when considering DA. On the one hand, we can standardize data characteristics by preprocessing, potentially reducing domain shift. We included two such steps by default: resampling to common spacing and intensity normalization; they are essential for the adequate model training \cite{kondrateva2024negligible}. Many studies demonstrated domain shift in medical images by intensity histograms \cite{crossmoda,se_medim,ihf}. Equalizing this difference might be of interest for adaptation, thus we included \textit{histogram matching (HM)}.

On the other hand, augmentations can expand source distribution, potentially covering the target one. Here, nnUnet framework \cite{nnunet} includes a variety of universal augmentations, so we tested them as a separate method under the name \textit{nnAugm}. We also tested a commonly used and modality-agnostic \textit{gamma correction augmentation (Gamma)} as an ablation study of nnAugm.
% as in \cite{gamma_example}

Finally, several advanced augmentation techniques were developed for domain generalization purposes. We included the most recent of them, \textit{global intensity non-linear (GIN)} \cite{gin} and \textit{modality independent neighborhood descriptor (MIND)} \cite{dg_tta} augmentations.

\textbf{Selected methods:} HM, nnAugm, Gamma, GIN, MIND.

\subsection{Implementation details}

\input{tables/suppl/4_hparams}

We used an nnU-Net \cite{nnunet} backbone as segmentation network architecture in all methods. We preserved most of the nnU-Net training pipeline except for several methodological changes, which allow us to evaluate DA methods, such as AdaBN and InstanceNorm, separately and run the ablation studies. These changes along with the other training hyper-parameters are summarized in Table~\ref{tab:hyper}.

Firstly, we replaced the default InstanceNorm with BatchNorm layers and removed test-time augmentation, so we can compare different normalizations and adaptive normalizations (AdaBN) and assess the unhindered impact of DA methods. Secondly, we reduced the patch size and number of the network features, so all experiments fit in a single 16 GB NVIDIA Tesla V100 and our benchmark remains economical. We set the number of epochs to 600 in all experiments, so that any method could complete its training in three days.

Below, we provide DA methods implementation details:

% \noindent
\textbf{Histogram matching} uses the baseline training pipeline, except all image intensity histograms are equalized to an average histogram computed over the train set.

% \noindent
\textbf{Gamma augmentation} also uses the baseline training pipeline, and we perform gamma correction with randomly selected $\gamma \sim U[0.5, 2]$ on every input image.

% \noindent
\textbf{nnAugm} similarly supplements the same baseline training with the original set of nnU-Net \cite{nnunet} augmentations.

% \noindent
\textbf{InstanceNorm} substitutes BN layers, while the training pipeline remains the same as in baseline.

% \noindent
\textbf{Adaptive BN} performs additional 1000 inference steps with batch size 4 over the baseline, updating the running statistics of BN layers on target training data.

% \noindent
\textbf{Self-ensembling} design and all parameters are reproduced from \cite{se_medim} with our architecture.

% \noindent
\textbf{MinEnt} adds a predictions entropy minimization criterion on target images. So we extended our training pipeline with the second step using target train images, and added entropy loss with the recommended in \cite{entropy} weight $\lambda = 0.001$.

% \noindent
\textbf{DANN} introduces an auxiliary network called discriminator. Similar to recent studies \cite{entropy}, we used DCGAN \cite{dcgan} discriminator architecture, replacing 2D convolutions with 3D ones. The losses weighting parameter is taken from \cite{dann_medim}, e.g., $\alpha = 0.01$.

% \noindent
\textbf{CycleGAN 2D} is fully reused from the original study \cite{cyclegan}. We trained a standalone CycleGAN 2D to map between source and target train images, where we sample axial slices from our volumetric images and rescale them into 256 $\times$ 256 gray scale images. Before predicting with the baseline segmentation model, we applied one of the generators to target test images (slice-by-slice) to transform them into fake source ones.

% \noindent
\textbf{CycleGAN 3D} is fully reused from the original study \cite{cyclegan3d}. We trained a standalone CycleGAN 3D to map between source and target train images, where we sample patches of size (128, 128, 96) from our volumetric images. Before predicting with the baseline segmentation model, we applied one of the generators to target test images (via overlapping grid) to transform them into fake source ones.

% \noindent
\textbf{GIN} is fully reused from \cite{gin} with the implementation based on the nnU-Net framework.

% \noindent
\textbf{MIND} is fully reused from \cite{dg_tta} with the implementation based on the nnUNet framework.

All experiments are available and could be reproduced from \href{https://github.com/BorisShirokikh/M3DA-exp}{https://github.com/BorisShirokikh/M3DA-exp}.

\begin{figure}
    \includegraphics[width=\linewidth]{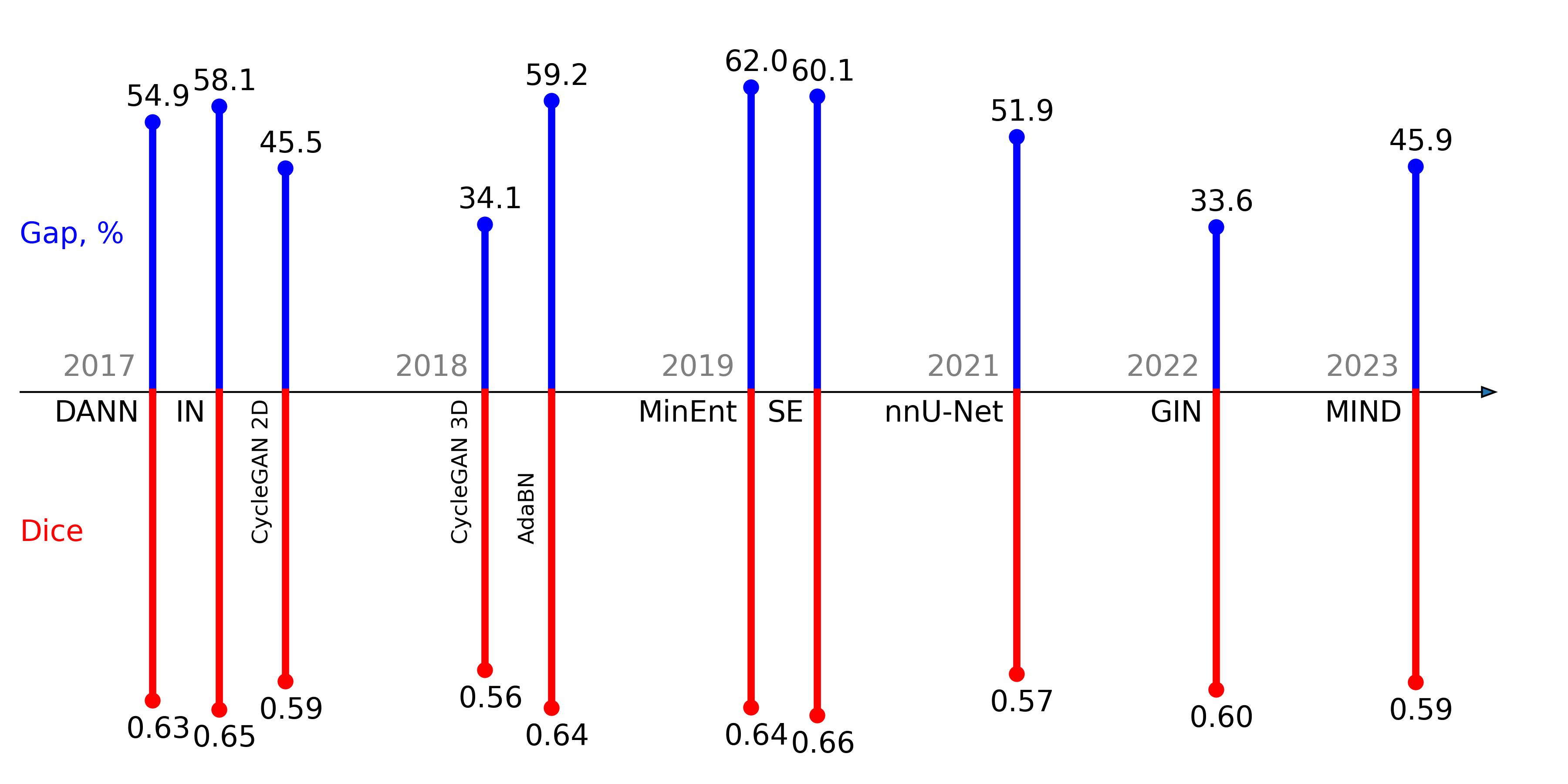}
    \caption{Average performance of domain adaptation approaches on M3DA benchmark; see Table \ref{tab:ablation_aug} for detailed results.}  % objective progress being very little over time
    \label{fig:teaser3}
\end{figure}

%% file: tables/suppl/4_hparams.tex
\begin{table}[h]
    \centering
    \caption{Hyper-parameters.}
    \label{tab:hyper}
    \resizebox{\columnwidth}{!}{
    \begin{tabular}{lcc}
        \toprule 
        \textbf{hyper-parameter} & \textbf{nnUNet} & \textbf{U-Net (Baseline)}  \\ 
        \midrule
        architecture & auto & auto \\
        base features & 32 & 24 \\
        normalization & instance (IN) & batch (BN) \\
        batch size & 2 & 2 \\
        patch size & (160, 192, 64) & (160, 160, 64) \\
        epochs & 600 & 600 \\
        batches per epoch & 250 & 250 \\
        loss & Dice Loss + CE & Dice Loss + CE \\
        % loss masking based on intensity & \cmark & \xmark \\
        oversampling rate & 0.66 & 0.75 \\
        optimizer & SGD & SGD \\
        momentum & 0.99 & 0.99 \\
        weight decay & $3 \times 10^{-5}$  & $3 \times 10^{-5}$ \\
        initial learning rate & $10^{-2}$ & $10^{-2}$ \\
        learning rate schedule & poly decay & poly decay \\
        learning rate decay power & 0.9 & 0.9 \\
        test-time augmentation & \cmark & \xmark \\
        \bottomrule

    \end{tabular}}
\end{table}

%% file: content/suppl/3_experiments.tex
\section{Supplementary experiments}

\begin{figure*}[h]
    \centering
    \includegraphics[width=0.85\linewidth]{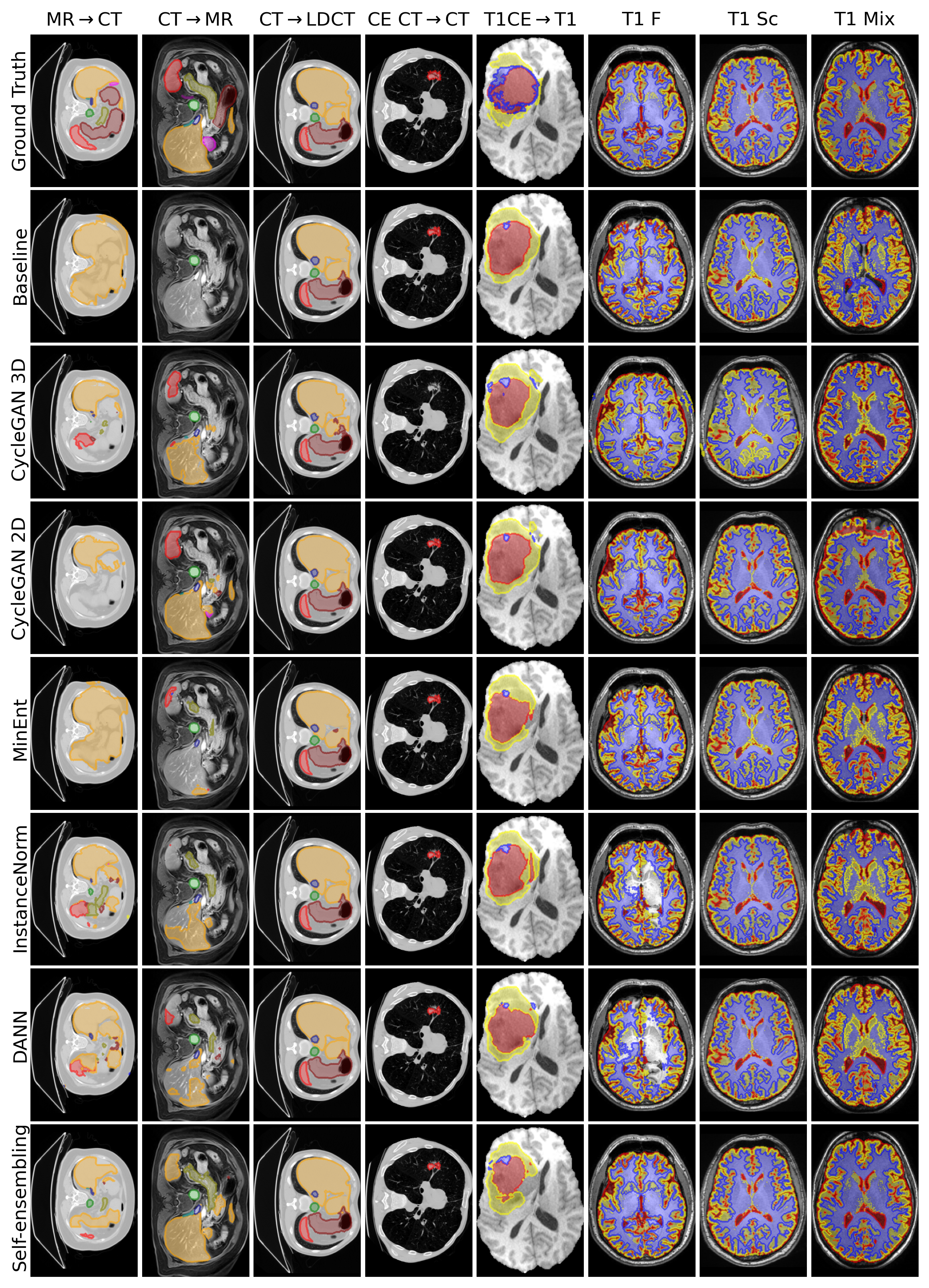}
    \caption{Example predictions for different DA methods. Methods are in rows, starting with ground truth in a first row. All eight benchmark tasks are in columns.}
    \label{fig:predicts}
\end{figure*}

\input{tables/suppl/5_t1t2_ct_kernels}

Firstly, Table~\ref{tab:t1t2_soft_hard} provides the results in four domain shift setups which we excluded from the benchmark. Differences in CT reconstruction kernels were excluded due to relative simplicity (Baseline is only marginally worse than the Oracle, and extensive augmentations almost closes this gap). We also excluded the domain shift between T1 and T2 MRI sequences, because, we were not able to find enough clinical justification for the inclusion.

Finally, example predictions for different DA methods are provided in Figure~\ref{fig:predicts}.

%% file: tables/suppl/5_t1t2_ct_kernels.tex
\begin{table}
    \centering
    \caption{Four setups not included in the M3DA benchmark.}

    \resizebox{0.5\textwidth}{!}{%
    \begin{tabular}{lcccc}
        \toprule
        & CT (soft) $\rightarrow$ CT (sharp) & CT (sharp) $\rightarrow$ CT (soft) & T1$\rightarrow$T2 & T2$\rightarrow$T1 \\
        
        \midrule
      
        \textit{Baseline} &  0.796	 &  0.819 &0.329 & 0.180  \\
        
        \midrule
        
        HM & --- & --- & 0.246 & 0.031 \\

        % CycleGAN 3D & --- & --- & --- & ---  \\

        MinEnt & --- & ---  & 0.322 & 0.094 \\
        
        CycleGAN 2D & --- & --- & 0.425 & 0.080  \\

        % GIN & --- & --- & --- & ---  \\

        AdaBN & --- & --- & 0.262 & 0.067 \\

        DANN &  --- & ---  & 0.228 & 0.028 \\

        IN &---	 & --- & 0.243 & 0.038 \\

        % MIND & --- & --- & --- & ---  \\

        Gamma & ---	 & ---	 & 0.288 & 0.132 \\

        SE & --- & --- & 0.000 & 0.000 \\
        
        nnAugm & 0.844 &	0.822 & 0.327 & 0.197 \\
        
        \midrule
        
        \textit{Oracle} &0.847 &0.845 & 0.728 & 0.686\\
        
        \bottomrule
         
    \end{tabular}}
    \label{tab:t1t2_soft_hard}
\end{table}